\PassOptionsToPackage{pdftex,
pdfencoding=auto,
pdfnewwindow=true,
pdfusetitle=true,
bookmarks=true,
bookmarksnumbered=true,
bookmarksopen=true,
pdfpagemode=UseThumbs,
bookmarksopenlevel=1,
pdfpagelabels=false,
breaklinks=true,
colorlinks=true
}{hyperref}
\PassOptionsToPackage{usenames,dvipsnames,pdftex}{xcolor}
\PassOptionsToPackage{letterpaper,margin=0cm,bottom=0.1cm}{geometry}
\documentclass[prx,11pt,letterpaper,twocolumn,accepted=2023-11-27]{quantumarticle}
\pdfoutput=1
\usepackage[letterpaper,margin=0cm,bottom=0.1cm]{geometry}
\usepackage[english]{babel}
\usepackage[numbers,sort&compress]{natbib}
\usepackage{amsmath,amssymb,graphicx,enumerate,bbm,xcolor,latexsym}
\usepackage{amsthm}
\usepackage{mathtools}
\usepackage{soul}
\usepackage{comment}
\usepackage[caption=false]{subfig}
\usepackage{paralist, tabularx}
\usepackage{nicefrac}
\usepackage[T1]{fontenc} 
\usepackage[utf8]{inputenx}

\usepackage[subfigure,titles]{tocloft}
\usepackage[page,title,toc]{appendix}

\usepackage{microtype}
\microtypecontext{spacing=nonfrench}
\microtypesetup{
expansion={true,nocompatibility},
protrusion={true,nocompatibility},
activate={true,nocompatibility},
tracking=true,
kerning=true,
spacing={true}
}
\usepackage[all=normal,floats=tight,mathspacing=tight,wordspacing=tight,paragraphs=normal,tracking=tight,charwidths=tight,mathdisplays=normal,sections=normal,margins=normal]{savetrees}
\everypar=\expandafter{\the\everypar\loosness=-1 }
\definecolor{myurlcolor}{rgb}{0,0,0.7}
\definecolor{myrefcolor}{rgb}{0.8,0,0}
\definecolor{purple}{RGB}{128,0,128}
\definecolor{ultramarine}{RGB}{63, 0, 255}
\definecolor{medblue}{RGB}{0, 0, 100}
\definecolor{googleblue}{RGB}{34, 0, 204}
\definecolor{panblue}{RGB}{0,24,150}
\definecolor{carmine}{RGB}{150, 0, 24}
\definecolor{gray}{RGB}{150, 150, 150}

\usepackage[unicode=true,pdfusetitle, bookmarks=false,bookmarksnumbered=false,
bookmarksopen=false, breaklinks=false,pdfborder={0 0 0},backref=false,
colorlinks=true,
linkcolor=carmine,
citecolor=googleblue,
urlcolor=panblue,
anchorcolor=OliveGreen,
]{hyperref}

\newtheorem{theorem}{Theorem}[section]

\newtheorem{defn}[theorem]{Definition}

\newcommand{\beq}{\begin{equation}}
\newcommand{\eeq}{\end{equation}}
\newcommand{\bea}{\begin{align}}
\newcommand{\eea}{\end{align}}
\newcommand{\bq}{\begin{quote}}
\newcommand{\eq}{\end{quote}}

\newcommand{\GPTt}{\ensuremath{\Omega}}
\newcommand{\gpEff}{\ensuremath{\mathcal{E}}}
\newcommand{\gp}[1]{\ensuremath{\mathbf{#1}}}
\usepackage[style=american]{csquotes}
\usepackage{centernot}

\usepackage{authblk}

\begin{document}
\title{A review and reformulation of macroscopic realism: \newline resolving its deficiencies using the framework of generalized probabilistic theories}

\author{David Schmid}
\email{david.schmid@ug.edu.pl}
\affiliation{International Centre for Theory of Quantum Technologies, University of Gdansk, 80-308 Gdansk, Poland}

\maketitle
\begin{abstract}
The notion of macrorealism was introduced by Leggett and Garg in an attempt to capture our intuitive conception of the macroscopic world, which seems difficult to reconcile with our knowledge of quantum physics. By now, numerous experimental witnesses have been proposed as methods of falsifying macrorealism. In this work, I critically review and analyze both the definition of macrorealism and the various proposed tests of it, identifying a number of problems with these (and revisiting key criticisms raised by other authors). I then show that all these problems can be resolved by reformulating macrorealism within the framework of generalized probabilistic theories. In particular, I argue that a theory should be considered to be macrorealist if and only if it describes every macroscopic system by a strictly classical (i.e., simplicial) generalized probabilistic theory. This approach brings significant clarity and precision to our understanding of macrorealism, and provides us with a host of new tools---both conceptual and technical---for studying macrorealism. I leverage this approach i) to clarify in what sense macrorealism is a notion of classicality, ii) to propose a new test of macrorealism that is maximally informative and theory-independent (unlike all prior tests of macrorealism), and iii) to show that every proof of generalized contextuality on a macroscopic system implies the failure of macrorealism.
\end{abstract}
\tableofcontents 

\section{Introduction}

A recurring theme in quantum foundations is the question of how to reconcile our quantum understanding of microscopic systems with our immediate experience of the macroscopic world. Why, for example, do we never observe interference with macroscopic objects? Is this just an experimental limitation due to environmental decoherence? Or might it be, as Bohr thought, that quantum theory is simply {\em wrong} at the macroscopic scale?

To explore this latter possibility, Leggett and Garg~\cite{LG1985} introduced macroscopic realism, or {\em macrorealism}, a notion of classicality which aimed to capture our experience of macroscopic objects such as chairs and tables---objects which appear to be in definite states at all times. They moreover proposed an experiment which aimed to be able to falsify macrorealism. 

Unfortunately, Leggett and Garg's definition of macrorealism is not precise by modern foundational standards. This point was made extensively and cogently by Timpson and Maroney~\cite{maroney2014quantum}, and I will revisit the matter herein.
Ultimately, I will argue that when one attempts to formalize macrorealism precisely, one is led to the following conclusion, which is the central thesis of this paper:
\begin{quote}
 {\bf Macrorealism is best characterized as the operational hypothesis that macroscopic systems are described by {\em strictly classical} generalized probabilistic theories.}
 \end{quote}
\noindent This unifies the notion of macrorealism with a well-studied class of theories  within the framework of generalized probabilistic theories~\cite{Hardy,GPT_Barrett,chiribella2010probabilistic} (GPTs)---roughly speaking, those whose state spaces are simplices and whose effect spaces are the dual hypercubes. (Such GPTs were introduced in Ref.~\cite{GPT_Barrett} and termed {\em strictly classical} in Ref.~\cite{selby2021accessible}.)

Moreover, the experimental proposals so far put forward for testing macrorealism have been quite controversial~\cite{maroney2014quantum,HardyLG,foster1991squid,Grassi1994,Clifton1991,GuidoLG}.
I will revisit and expand on the most important of these criticisms in order to motivate the need for better tests of macrorealism. I will then argue that, using the above characterization of macrorealism within the framework of generalized probabilistic theories, one naturally reaches the second key conclusion of this paper:
\begin{quote}
{\bf The ideal way to test macrorealism is to use theory-agnostic GPT tomography~\cite{bootstrap1,bootstrap2}, which makes significantly weaker assumptions than prior tests, while yielding a complete characterization of one's system (rather than just a witness of the failure of macrorealism).}
\end{quote}

A third aim of this paper is to provide a concise and accessible review of the literature on macroscopic realism, from the original work~\cite{LG1985} through the major subsequent criticisms and proposed revisions to which this has been subjected~\cite{maroney2014quantum,HardyLG,foster1991squid,Grassi1994,Clifton1991,GuidoLG,KoflerBrukner,knee2016strict,wilde2012addressing,allen2017stronger}. These criticisms and proposed developments help to motivate the value of the two proposals above in bold.

Regarding the definition of macrorealism, an insightful analysis was given by Timpson and Maroney~\cite{maroney2014quantum}. There, the authors begin by extensively arguing that the traditional definition of macrorealism is ambiguous (primarily because it makes reference to an undefined notion of state) and does not pick out a particularly natural class of theories. 
Timpson and Maroney then suggest a significantly more precise version of macrorealism, which they termed {\em operational eigenstate mixture macrorealism}. (This improved definition was indeed later adopted and used to great effect by Leggett and coauthors~\cite{knee2016strict}.)
In this work, I cast Timpson and Maroney's definition into the framework of generalized probabilistic theories.  This reframing unifies macrorealism with the notion of strict classicality~\cite{GPT_Barrett} that has been independently motivated and studied within the framework of GPTs. This brings further precision and clarity to our conceptual understanding of macrorealism, and moreover extends it into a full-bodied notion of a theory (not merely a statement about the possible states in the theory). Moreover, viewing macrorealism within the GPT framework implies that one can bring all the well-developed tools and concepts within this framework to bear on the problem of testing macrorealism.

Regarding the problem of testing macrorealism, I will first analyze three types of prior proposals for testing macrorealism: tests based on Leggett-Garg inequalities~\cite{LG1985}, on no-signaling in time~\cite{KoflerBrukner}, and on nondisturbance conditions~\cite{knee2016strict}. The first two types of tests rely on the assumption that one can implement measurements in a noninvasive manner, and I will revisit and expand on prior criticisms~\cite{maroney2014quantum,HardyLG,foster1991squid,Grassi1994,Clifton1991,GuidoLG} of this assumption. In particular, I discuss how one can only justify noninvasiveness by making detailed theory-dependent assumptions about the nature of the system in question and the nature of the measurements used to probe it. In contrast, the third type of test~\cite{knee2016strict} avoids any need for such assumptions regarding the nature of one's measurements. However, I will show that these tests rely on an unphysical idealization regarding the {\em preparations} used in one's experiment. Moreover, the only way to make the experiment robust to relaxations of these idealizations is to introduce theory-dependent assumptions much like those required in the first two types of tests. Consequently, tests of this third sort fare no better than prior tests in terms of achieving theory-independence (although they certainly are more  elegant and generally applicable than tests based on Leggett-Garg inequalities). 

Thankfully, the reframing of macrorealism as strict classicality in the framework of GPTs suggests a better approach to testing macrorealism---one which makes no a priori assumptions about the nature of the states and measurements one uses to probe the system in question.
In particular, a technique known as theory-agnostic tomography~\cite{bootstrap1,bootstrap2} (or bootstrap tomography) allows one to {\em extract} this information---a complete characterization of the GPT descriptions of every state and effect realized in one's experiment, and consequently an (inner) approximation of the true GPT state space. As I elaborate below, this provides maximal information about the nature of the macroscopic system being probed, and does so with significantly fewer assumptions than all prior tests of macrorealism.

Finally, I leverage the GPT characterization of macrorealism to clarify several of its features. First, I expand on the fact (also argued by Timpson and Maroney~\cite{maroney2014quantum}) that macrorealism is not a notion of realism---rather, it is a (very strict) notion of classicality. Then, I contrast this notion of classicality with the much more permissive notion of classicality given by generalized noncontextuality~\cite{gencontext}. In particular, I show that macrorealism implies that every macroscopic system satisfies generalized noncontextuality, and hence that any proof of the failure of generalized noncontextuality for a macroscopic system implies the failure of macrorealism.

\section{The traditional definition of macrorealism}

The standard definition of macrorealism that one finds in the literature is that given by Leggett and Garg~\cite{LG1985,leggett2002testing}.
\begin{defn}
Macrorealism is defined by the conjunction of two\footnote{The implicit assumption of no-retrocausation is sometimes explicitly included as a third assumption (and is often termed `induction').} assumptions.
\begin{itemize} \label{LGdefn}
{\em \item{\em Macrorealism per se.} A macroscopic object which has available to it two or more macroscopically distinct states is at any given time in a definite one of those states.
\item{\em Noninvasive measurability (at the macro level).} It is possible in principle to determine which of these states the system is in without any effect on the state itself or on the subsequent system dynamics. }
\end{itemize}
\end{defn}

As evidenced by its name, the first of these assumptions is the essence of what is meant by macroscopic realism. It aims to capture the {\em ``whole class of theories or models which have in common only the property that quantum superpositions of macroscopically distinct states do not exist in nature; exactly at what stage and by what mechanism such superpositions are avoided is left open''.}\cite{leggett2002testing}

Leggett argues that the second assumption, noninvasive measurability, is {\em ``so natural a corollary of macrorealism per se that the latter is virtually meaningless in its absence''}~\cite{leggett2002testing}. However, others have argued that this assumption is not necessary for a theory to be reasonably called realist about the macroscopic. While I will ultimately argue that noninvasive measurability is indeed a natural assumption to consider in conjunction with macrorealism per se, {\em experimental tests} of macrorealism which rely on it unfortunately {\em also} rely on a stronger assumption, namely
\begin{itemize}
\item{\em Realized noninvasiveness.} The particular measurement that is actually implemented in one's experiment does not have any effect on the state itself or on the subsequent system dynamics.
\end{itemize}
Note that while noninvasive measurability is a property of a theory, realized noninvasiveness is an assumption about a particular experiment. Clearly, realized noninvasiveness can only be satisfied in theories satisfying noninvasive measurability. (In most discussions of macrorealism, noninvasive measurability and realized noninvasiveness are conflated. However, it will be useful to separate the two.)

Let us now analyze each of these assumptions in more detail.

\subsection{Macrorealism per se} \label{mrps}

The assumption of macrorealism per se, as stated, has two notable ambiguities. 

The first of these is that it refers to an undefined notion of macroscopicity. Defining macroscopicity is a difficult and interesting problem in its own right~\cite{RevModPhys.90.025004}. This ambiguity is no problem at all, however---Leggett and Garg's reasoning can be applied to any notion of macroscopicity that one cares to consider. Note, however, that experiments aiming to witness the failure of macrorealism can only be considered to have foundational significance if the system of interest is deemed macroscopic by a definition that one considers to be compelling.

The second ambiguity is more serious, and arises from the fact that the notion of a `state' is dependent on the theory or framework one has in mind. In quantum theory, the notion of state is the density operator, and in the framework of generalized probabilistic theories~\cite{Hardy,GPT_Barrett,chiribella2010probabilistic}, the notion of state is a list of probabilities for a set of fiducial measurements. In a classical realist framework~\cite{schmid2020unscrambling} (e.g., the ontological models framework~\cite{Harrigan}), there are two notions of state: the ontic state (the fundamental description of the properties of a system) and the epistemic state (an agent's knowledge of the ontic state).
In still other frameworks, the notion of state may be neither of these.
Which notion of state one has in mind has significant consequences for one's understanding of Leggett and Garg's assumption. 

Timpson and Maroney discuss this ambiguity in detail in Ref.~\cite{maroney2014quantum}, where they argue that the background framework in which Leggett and Garg frame their arguments and definitions is simply quantum mechanical Hilbert space states:
\begin{quote}
{\em Macroscopic states are the quantum states that one would assign to macroscopic, or collective, degrees of freedom.} Thus, in a SQUID, one does not trouble to assign a (massively entangled) multi-particle quantum state to the {\em enormous} number of individual microscopic charge-carriers, rather one simply assigns a single state to the collective degree of freedom, the direction of the current, e.g., $|+1\rangle$ or $|-1\rangle$. The content of macroscopic realism is then that the only permissible states of the SQUID are the quantum states $|+1\rangle$ and $|-1\rangle$ (and statistical mixtures thereof), quantum superpositions of these two states being disallowed.
\end{quote}
They support this reading with an analysis of Leggett's writing\footnote{Furthermore, one can find clear textual evidence that Leggett and Garg do {\em not} intend the word state to refer to an ontic state in this context. As one example, consider the final phrase of this description of the noninvasive measurability assumption: ``{\em It is possible, at least in principle, to measure the quantity $Q(t)$ at a particular time $t$ in such a way that the subsequent dynamics of the system is unaffected, at least as regards its motion between the macrostates of interest.}''~\cite{leggett1988experimental}}.
In a later paper~\cite{knee2016strict}, Leggett and coauthors are explicit that this is a fair reading of their assumptions: they denote the off-diagonal terms in a standard quantum density operator by $\mathcal{C}$, and then state {\em ``In this language the predictions of a macrorealist theory (for super-critically macroscopic systems) are equivalent to those which follow from putting $\mathcal{C}=0$.''}
So, in Timpson and Maroney's words, 
\begin{quote}
It seems that we are in the realm of applying quantum states to describe physical systems, it’s just that not all the usual states in the Hilbert space are permitted as physical. (It may be that we do not always get a new physical state by adding together two of the old ones; but the old ones—the items one is contemplating adding together—are themselves quantum states.)
\end{quote}

In other words, Leggett and Garg's assumption of macrorealism essentially postulates that there is a superselection rule~\cite{bartlett2007reference} which forbids macroscopic systems from having coherence with respect to some privileged basis. In other words, the state of such a system is always diagonal in this basis and processes preserve the basis. Any such system is classical in the strictest sense, and can be described simply as a classical random variable. So we see that Leggett and Garg's real aim is simply to determine whether macroscopic systems in our world are adequately described by random variables. Somehow, this simple fact seems to have been masked by the vagueness in Leggett and Garg's choice of definitions and introduction of the novel term ``macrorealist'' where the preexisting language of superselection rules or random variables would have sufficed.\footnote{In fairness to Leggett and Garg, at the time of their work there was no well-developed framework for referring to alternative ways the world could be---like the GPT framework---which might be why they struggled to define a notion like `diagonal quantum theory' without referencing quantum theory itself.}  Moreover, macrorealism has very little to do with realism, but rather is a notion of classicality, as I will discuss in Section~\ref{realism}. So it is unfortunate that these definitions and terminology have been adopted rather than improved in most subsequent work. 

One important exception, however, is the work of Timpson and Maroney~\cite{maroney2014quantum}, which did a good deal to clarify the relevant concepts and terminology. Timpson and Maroney reframe the discussion of macrorealism in terms of operational theories\footnote{Note that they primarily consider unquotiented operational theories, whereas I will work herein with quotiented operational theories, according to the distinction defined in Refs.~\cite{chiribella2010probabilistic,schmid2020structure}.} and ontological models~\cite{Harrigan}. They then propose a more precise definition of macrorealism\footnote{In fact, they introduce three different notions of macrorealism, only the strictest of which is directly relevant to this work. (These three notions share the essential feature that {\em macroscopic observables in the theory should be assigned definite and noncontextual values}. This is such a broad notion of macrorealism that it can be consistent with all of quantum theory. For example, if one's only macroscopic quantities are (possibly coarse-grained) position observables, then Bohmian mechanics satisfies this criteria~\cite{maroney2014quantum}. }, based on the notion of an {\em operational eigenstate}---any preparation of a system which gives deterministic outcomes for the macroscopic observable. (This is simply an operational generalization of a quantum eigenstate.) 
\begin{defn}\label{opeigendefn}
A theory satisfies {\em operational eigenstate mixture macrorealism} if the only possible preparation states of a macroscopic system are operational eigenstates of the given macroscopic observable and statistical mixtures thereof. 
\end{defn}

This definition captures the case of quantum theory with a macroscopic superselection rule, and seems to capture (and make precise) Leggett and Garg's notion of macrorealism. Indeed, Leggett and coauthors endorse this  definition in Ref.~\cite{knee2016strict}.
It seems this definition moreover captures most of the views on macrorealism in the literature.\footnote{Timpson and Maroney's two more permissive versions of macrorealism are the only explicit exceptions I am aware of.}

An interesting feature of Definition~\ref{opeigendefn} is that this notion of macrorealism is phrased entirely in terms of operationally accessible concepts. No reference is made to concepts such as ontic states or causal structure (which are not always operationally accessible). This feature is {\em not} shared by the notion of realism that is required by Bell. It is, however, consistent with the reading of macrorealism as a strict notion of classicality---namely, that a system can be completely described by a classical random variable. 

This suggests that the notion of macrorealism can be characterized within the framework of generalized probabilistic theories. In Section~\ref{betterdefn}, I discuss how this recasting of the definition leads directly to a previously studied notion of strictly classical GPTs~\cite{GPT_Barrett,selby2021accessible}, and I argue that this characterization of macrorealism has a number of additional benefits.

\subsection{Noninvasive measurability}

In most discussions of macrorealism, noninvasive measurability and realized noninvasiveness are conflated. However, these two assumptions are entirely different, both as a matter of logic, and in terms of their overall plausibility and testability. 
Moreover, many of the criticisms that have been levelled at noninvasive measurability are better understood as criticisms of realized noninvasiveness.
I will come to these criticisms in the next section, where realized noninvasiveness is discussed.

Here, we are simply concerned with the question: is noninvasive measurability of macroscopic systems a reasonable constraint on any theory deserving to be called `realist about the macroscopic'?

Certainly, noninvasive measurability should not be described as a `corollary' of macrorealism per se, despite Leggett's penchant for saying just this. Noninvasive measurability simply is {\em not} a logical corollary of macrorealism per se. (If it were, there would be no need to include it as an assumption in discussions of macrorealism! So even Leggett himself clearly does not view it as a logical corollary.) The assumption of macrorealism per se says nothing to constrain the sorts of measurements that are possible in one's theory. 

Many commentators~\cite{maroney2014quantum,HardyLG,foster1991squid,Grassi1994,Clifton1991,GuidoLG} have argued that there exist theories wherein noninvasive measurability does not hold, and yet which are reasonably said to be realist about the macroscopic. For instance, Timpson and Maroney consider a broader notion of macrorealism (which states that macroscopic observables in the theory are always assigned definite and noncontextual values), and argue that {\em ``...noninvasiveness cannot be motivated as a general feature which should exist for at least some measurements when macroscopic realism obtains.''}~\cite{maroney2014quantum} For instance, they note that Bohmian mechanics satisfies this broader notion of macrorealism relative to position observables, and yet measurements on a system certainly disturb the later dynamics of the particle's position (by virtue of disturbing the wavefunction)~\cite{maroney2014quantum,GuidoLG}. Similarly, the toy field theory in Ref.~\cite{catani2021interference} satisfies this notion of macrorealism relative to occupation number and phase observables, and yet the theory does not allow measurements of either of these that are  nondisturbing.

However, one certainly {\em can} define classes of theories that include noninvasive measurability as an assumption, and it is this class of theories that macrorealism has historically been about.
Moreover, this class of theories {\em is} a natural one to consider, {\em provided} one is interested in singling out the strictest notion of classicality---e.g., the one which seems to describe our experience of, say, tables and chairs. Clearly, such macroscopic objects can be observed in a manner that does not disturb their macroscopic state, and it is this intuition that motivated the study of macrorealism in the first place~\cite{LG1985}. 

Moreover, the assumption of noninvasive measurability {\em does} naturally form a part of the strictest notion of classicality that is typically considered in the framework of GPTs, as I will discuss further in Section~\ref{simpliciality}.

One likely reason that so many foundational researchers have resisted the assumption of noninvasiveness is that foundational arguments typically aim to invoke as weak assumptions as possible, so that one's arguments rule out as broad a class of physical theories as possible. But Leggett and Garg's work appears to have a very different kind of aim---it aims to characterize and rule out a {\em maximally stringent} notion of classicality. (I return to the question of whether this is as well-motivated as the study of more permissive notions of classicality in Section~\ref{MRvsNC}.)

\subsection{Realized noninvasiveness} \label{realizedNIM}

We turn now to the more problematic assumption of realized noninvasiveness. The basic problem is the following: how can one ever ensure that {\em the particular measurement used in one's experiment} is minimally invasive (and hence, by noninvasive measurability, totally noninvasive)? 
This assumption has been criticized in many prior works, e.g. Refs.~\cite{maroney2014quantum,HardyLG,foster1991squid,Grassi1994,Clifton1991,GuidoLG}, and the possibility that one's measurements are not minimally invasive is often termed the clumsiness loophole.
 
 A priori, it would seem that experimental imperfections in any real measurement (even of a macroscopic observable) will cause a nonzero disturbance on the system being measured. In response to this concern, Leggett and Garg argue that {\em under the assumption of macrorealism}, one can in fact implement a perfectly nondisturbing measurement by carrying out a procedure known as a null-result measurement. (Note that this is usually given as an argument for noninvasive measurability, but it implies both noninvasive measurability {\em and} the stronger assumption of realized noninvasiveness.)

A null result experiment is defined as one in which the measuring apparatus interacts with the system {\em only} if the system is in a particular operational eigenstate, but does not interact with the system if it is in any other operational eigenstate. Under the assumption of macrorealism per se, the state {\em is} always an operational eigenstate or a convex mixture thereof, and so one need not specify whether or not there is an interaction for any other possible ways of preparing the system.

To illustrate the argument, Leggett often gives an example of what he considers to be a null-result measurement~\cite{LG1985,leggett1988experimental}---an example which has also been studied under the alternative name `interaction-free' measurement~\cite{elitzur1993quantum}. The example is a particular implementation of a measurement of which path a particle took in a double-slit experiment. One simply places an absorbing particle detector in the lower of the two slits, and then post-selects on cases where this detector does not fire. Under the assumption that the which-way observable is macrorealist, the particle travels through {\em either} the upper slit {\em or} the lower slit, and so one can infer that the particle traveled through the {\em upper} slit from the fact that the detector did not fire. Hence, Leggett argues, the detector cannot have influenced the particle, as the two are in separate locations at all times. 
So, one is meant to start from the assumption of macrorealism, and is meant to {\em infer} realized noninvasiveness for this measurement procedure (and consequently noninvasive measurability as well).\footnote{In quantum theory, of course, the interference pattern always disappears when the which-path information is measured in any way, even via such a null-result measurement. This by itself does not falsify the argument, however, since quantum particles do not satisfy the assumption of macrorealism used in the argument.} 

However, this argument has implicitly relied on a number of extra assumptions. For one, it clearly relies on an assumption that causal influences are local. Moreover, it relies on an assumption that there are exactly two possible properties that the system may have: `located at the right slit and `located at the left slit'. This is not as innocent an assumption as it first appears. The issue is not that one is assuming that the particle has a definite location, for this is part and parcel of macrorealism. Rather, it is that one is assuming {\em that there are no {\em other} properties of a particle that could possibly be relevant to its behavior in a double slit experiment}. It is unclear what would motivate such an assumption, especially once one observes that there are operations (like a phase shifter at one slit) which affect the overall observed behavior of the particle while not affecting the particle's location. 

If one allows the possibility that physical properties {\em other} than particle position which are relevant to explaining interference phenomena, then one can find explicit models which are macrorealist and yet where implementing a null-result measurement does {\em not} imply realized noninvasiveness. 
One such model can be constructed from the toy field theory in Ref.~\cite{catani2021interference}, which models interference in a two-mode interferometer.  A key part of the model is to treat each mode of an interferometer as a physical system that contains not only a property corresponding to particle number, but also a property corresponding to phase, contradicting the assumption I identified above (that there are only two possible states of the system in question). In this model, placing a detector at the lower path does not influence the position of the particle when the particle takes the upper path, just as Leggett and Garg's argument suggests. {\em However}, the measurement nonetheless {\em does} (locally) disturb the {\em phase} of the lower mode, which consequently can influence the particle's dynamics at a later time. So this model demonstrates that the implication from null result measurements to realized noninvasiveness does indeed rely on the implicit assumption above\footnote{Strictly speaking, this only constitutes a counterexample if the model in question is macrorealist. 
While the model as presented in Ref.~\cite{catani2021interference} is not macrorealist in the sense of strict classicality that I argue for below, it can easily be modified to be (without changing the statistics it can predict and without changing the argument above.) This follows immediately from the fact that the model is noncontextual, and hence simplex-embeddable, and so all the statistics it predicts can be reproduced in a strictly classical (simplicial) GPT.}. Indeed, one of the conclusions of Ref.~\cite{catani2021interference} is that, contrary to the intuitions expressed in Refs.~\cite{LG1985,leggett1988experimental,elitzur1993quantum}, such measurements cannot be said to be interaction-free (at least without further assumptions, such as the assumption of psi-onticity---that the quantum state corresponds to an element of reality, rather than a representation of knowledge about reality~\cite{Harrigan,caves2002quantum,spekkens2007evidence}). 
 
Of course, one might reply that macrorealism is often {\em defined} with respect to some given quantum observable (as it is in Timpson-Maroney's definition, for example), and this fixes the number of operational eigenstates and the dimensionality of the system in question. If one grants that the dimensionality of the system is two, then {\em surely} the two associated operational eigenstates correspond to the particle being at one slit or the other.
But the question is {\em why} this should be granted, and {\em why} macrorealism should be defined with respect to any quantum observable that is given a priori---especially as we are in the business of imagining the possibility that quantum theory is not the correct description of nature. Better would be to presume nothing a priori about the nature or number of the operational eigenstates, and instead to learn what these are by experiment. Beginning with a particular quantum observable (or assuming a particular set of operational eigenstates) constitutes another implicit theory-dependent assumption going into standard analyses of macrorealism. If one gives up on this assumption, then one can no longer justify realized noninvasiveness or noninvasive measurability via null-result measurements. Moreover, we will see in Section~\ref{error} that this same implicit assumption is also needed (albeit for different reasons) in more recent tests of macrorealism that do not assume realized noninvasiveness or noninvasive measurability~\cite{knee2016strict}. 

In conclusion, even if one grants that it is sensible to include noninvasive measurability in one's definition of macrorealism (as I have argued one should), the stronger assumption of realized noninvasiveness is more problematic. Prior attempts to justify it have relied on additional assumptions that are easily challenged---most seriously, theory-dependent and model-dependent assumptions about the number and nature of possible physical properties of the system in question.

Whether or not one can provide arguments for these assumptions, tests which rely on them are made weaker for it. Any such argument rules out {\em not} the full class of macrorealist theories, but only those macrorealist theories satisfying the extra assumptions.
As Timpson and Maroney put it:
\begin{quote}
If one happens to have certain views as to how the detailed physics of the interaction between system and measuring apparatus goes, then one might very well believe that one had a pair of measurements apt for null-result, ontically noninvasive, measurement. But it could well be that one’s model is wrong, rather than that it is macroscopic realism which is at fault.''\cite{maroney2014quantum}
\end{quote}
Consequently, prior proposals for testing macrorealism are not 
methodologically on par with state-of-the-art foundational experiments such as Bell tests~\cite{Belltest1,PhysRevLett.115.250402,PhysRevLett.115.250401} or tests of generalized noncontextuality~\cite{Mazurek2016,operationalks}.

\section{A better definition of macrorealism}\label{betterdefn}

I now provide an alternative characterization of macrorealism that subsumes the traditional definition, but that I believe is more transparent and precise. This is accomplished using the framework of generalized probabilistic theories. We first take a quick detour to introduce the basics of this framework, following Refs.~\cite{GPT_Barrett,SchmidGPT}.

\subsection{Generalized probabilistic theories}
The framework of generalized probabilistic theories~\cite{Hardy,GPT_Barrett,chiribella2010probabilistic}, or GPTs, describes a landscape of potential theories of the world, with classical and quantum theories as special cases. Each GPT is characterized (solely) by the operational statistics it predicts. 

A system in a given GPT is associated with a convex set of normalized states (including mixed states),  $\Omega$, standardly assumed to be finite dimensional and compact. A well-known example is the Bloch ball in quantum theory, which can be viewed as the GPT state space of a qubit. While $\Omega$ naturally lives inside an affine space $\mathsf{AffSpan}[\Omega]$, it is often represented as living in a Euclidean vector space $V$ of one dimension higher, so that $\mathsf{AffSpan}[\Omega]$ is embedded as a hyperplane in $V$ which does not intersect with the origin $\boldsymbol{0}$. (This is analogous to embedding the Bloch ball within the real vector space of Hermitian matrices.) This allows us to define both GPT states and GPT effects (defined below) within the same vector space. The set of subnormalized states is then given by the convex hull of $\Omega$ and the zero vector $\boldsymbol{0}$.

Every system in a GPT is also associated with a set of effect vectors, $\mathcal{E}$, where the probability of obtaining an effect $\boldsymbol{e}\in\mathcal{E}$
given a state $\boldsymbol{s}\in\Omega$ is given by the dot product:
\beq
\mathrm{Prob}(\boldsymbol{e},\boldsymbol{s}) := \boldsymbol{e} \cdot \boldsymbol{s}.
\eeq
The set $\mathcal{E}$ is required to satisfy the following constraints.
Define the dual of $\Omega$, denoted $\Omega^*$, as the set of vectors in $V$ whose inner product with all state vectors in $\GPTt$ is between $0$ and $1$:
\beq
\Omega^* := \{ \boldsymbol{x}\in V | \boldsymbol{x}\cdot\boldsymbol{s} \in [0,1] \ \forall \boldsymbol{s} \in \Omega\}.
\eeq
Then, $\mathcal{E}$ must be a compact convex set contained in $\Omega^*$, $\mathcal{E} \subseteq \Omega^*$,
which contains the zero vector $\boldsymbol{0}$ and the unit effect $\boldsymbol{u}$, defined, respectively, by $\boldsymbol{0} \cdot \boldsymbol{s} = 0$ and $\boldsymbol{u} \cdot \boldsymbol{s} = 1$ for all $\boldsymbol{s}\in\Omega$.
Due to the choice of embedding of $\mathsf{AffSpan}[\Omega]$ within $V$, $\boldsymbol{u}$ necessarily exists and is unique~\cite{chiribella2010probabilistic}. Finally, $\mathcal{E}$ has the property that for all ${\bf e} \in \mathcal{E}$, one has ${\bf u} - {\bf e} \in \mathcal{E}$.

Moreover, the sets $\Omega$ and $\mathcal{E}$ in any valid GPT must satisfy the principle of {\em tomography}. This is the requirement that a GPT state can be uniquely identified by the probabilities it generates on GPT effects, so that $\gp{e}\cdot \gp{s_1} = \gp{e}\cdot\gp{s_2}$ for all $\gp{e}\in\gpEff$ if and only if $\gp{s_1} = \gp{s_2}$, and analogously, that a GPT effect can be uniquely identified by the probabilities it generates on GPT states, so that $\gp{e_1}\cdot\gp{s} = \gp{e_2}\cdot \gp{s}$ for all $\gp{s}\in\GPTt$ if and only if $\gp{e_1} = \gp{e_2}$.

The set $\mathcal{T}$ of normalization-preserving transformations from a GPT system to itself is given by a convex, compact set of linear maps\footnote{In the case of GPTs that do not satisfy tomographic locality~\cite{Hardy,hardy2012limited}, these linear maps only describe the action of the transformations on single systems; there may be many distinct transformations that induce the same linear map on a single system. This subtlety will not matter here, as we will focus on single systems. } from $V$ to itself, where each linear map takes elements of $\Omega$ to elements of $\Omega$, and its adjoint takes elements of $\mathcal{E}$ to elements of $\mathcal{E}$. The set of transformations must be closed under composition and include the identity transformation. Additionally, valid transformations must satisfy the analogue of complete-positivity~\cite{nielsen2001quantum,Schmidcausal}. 
One can also define the set $\mathcal{U}$ of normalization-nonincreasing transformations, which includes the convex hull of $\mathcal{T}$ together with the transformation that maps all states to the origin $\boldsymbol{0}$, and has the property that for each element of $\mathcal{U}$, there exists another element in $\mathcal{U}$ such that the two sum to an element in $\mathcal{T}$.
(One can in fact obtain further constraints by considering GPTs as compositional theories~\cite{hardy2011reformulating,chiribella2010probabilistic}; for instance, the set of transformations must include the measure-and-reprepare channels. These constraints will not be needed here.)

Finally, we define the set $\mathcal{I}$ of {\em instruments}; that is, nondestructive measurements on a system. In quantum theory, an instrument corresponds to a collection of completely-positive and trace-nonincreasing maps that sum to a completely positive trace-preserving map. Analogously, an instrument in a GPT is a set of normalization-nonincreasing transformations (elements of $\mathcal{U}$) that sum to a normalization-preserving transformation (an element of $\mathcal{T}$). 
The set of destructive measurements on a system, of course, can be defined simply by ignoring the output system in a nondestructive measurement. That is, one obtains a measurement by composing each element in a given instrument with the unit effect. The set of all measurements can be obtained from $\mathcal{I}$ in this manner.

In summary, the GPT representation of a single system is defined by a tuple $G:=(V, \Omega, \mathcal{E}, \mathcal{T}, \mathcal{U}, \mathcal{I})$ satisfying the above constraints.\footnote{ Note that the elements of this tuple are not all independent sets; for instance, $\mathcal{E}$ can be derived from $\mathcal{I}$.}

\subsection{Strictly classical GPTs} \label{simpliciality}
Within the GPT framework, there is a standard notion of a strictly classical theory~\cite{GPT_Barrett}. (The modifier `strictly' was first added in Ref.~\cite{selby2021accessible}, because it is also useful to consider a more permissive notion of classicality, namely generalized noncontextuality, i.e., simplex-embeddability of one's GPT, as I discuss in Section~\ref{MRvsNC}.) 

A strictly classical system of dimension $d$ can be defined as follows. Consider a system which is always in one of $d$ possible states, and for which there exists a single measurement that can exactly determine which state the system is in, and moreover can do so without disturbance. These $d$ states are represented by a set of $d$ linearly independent vectors living in some $d$-dimensional Euclidean vector space $V_d$, so that the full GPT state space is given by the convex hull of these, forming a simplex that I denote $\Delta_{d}$. Each point in the simplex is in one-to-one correspondence with a probability distribution over the true states of the system. The existence of the perfectly discriminating measurement in the theory implies that every logically possible effect is physically possible--- that is, the set of effects is taken to be the dual of $\Delta_{d}$, which I denote $\Delta_{d}^*$. This set of effects forms a hypercube. (If one chooses the representation of the extremal GPT states to be length-$d$ unit vectors, for example, then the vertices of the hypercube are all and only the length-$d$ vectors with all components equal to $0$ and $1$.) The allowed transformations in the strictly classical theory are the linear maps induced by functions from the vertices to the vertices and convex combinations thereof. This set includes all logically possible stochastic maps from a $d$-element set to a $d$-element set, and I denote it $\mathcal{T}_d$. The set of normalization-nonincreasing transformations is simply the set of substochastic maps of this type, and I denote it $\mathcal{U}_d$. Finally, an instrument is a set of such substochastic maps that sums to a stochastic map, and the set of all instruments is given by the set of all logically possible instruments and denoted $\mathcal{I}_d$. (This can in fact be derived by considering all possible postprocessings of the perfectly discriminating and nondisturbing measurement contained in the theory. It moreover follows that all measurements in the theory are compatible.) 

\begin{defn}{Strictly classical GPT.}
A strictly classical GPT of given dimension $d$ is a tuple  $(V_d, \Delta_{d},\Delta_{d}^*,\mathcal{T}_d,\mathcal{U}_d,\mathcal{I}_d)$ defined as above. 
\end{defn}
In the context of prepare-measure scenarios, such GPTs are often referred to as simplicial GPTs, in reference to the shape of the state space. 

It should be clear by their construction that strictly classical GPTs are simply an informational description of classical physics---that is, of systems described by classical random variables and which evolve by stochastic dynamics. 
Indeed, in a strictly simplicial theory, an arbitrary process (with any number of input systems, output systems, settings, and outcomes) is described by a substochastic map from the random variables associated with the input systems and settings to the random variables associated with the output systems and outcomes; moreover, {\em all} such substochastic maps are valid processes in the theory.

\subsection{Macrorealism as strict classicality} \label{simpliciality}

Following from my arguments in Section~\ref{mrps}, I propose the following as an improved definition of macrorealism.
\begin{defn} \label{mydefn}
{\bf A theory is {\em macrorealist} if and only if it describes every macroscopic degree of freedom by a strictly classical GPT.} 
\end{defn}
\noindent Similarly, I will say that a macroscopic degree of freedom\footnote{Throughout, I take the term `degree of freedom' to be synonymous with `system'. In the framework of GPTs, systems are primitive notions.} is macrorealist if it is described by a strictly classical GPT. 

Leggett and Garg's assumption of macrorealism per se is here formalized as the statement that the GPT state space is a simplex, where the `macroscopically distinct states' in question correspond to the vertices of the simplex, and the points inside the simplex represent states of uncertainty about which of these states the system is in. Moreover, Leggett and Garg's notion of noninvasive measurability is here formalized by the fact that the strictly simplicial theory includes a nondisturbing and maximally informative measurement. The existence of this measurement moreover ensures that the full hypercube of effects that is dual to the simplicial state space is included in the theory. 

This definition subsumes the notions of macrorealism I discussed above, namely  quantum systems under a superselection rule, as well as Timpson and Maroney's notion of operational eigenstate mixture macrorealism~\cite{maroney2014quantum} (which Leggett later endorsed~\cite{knee2016strict}). Hence, this definition can essentially be viewed as the characterization of the standard notion of macrorealism within the framework of GPTs. For this reason, I will use the terms {\em strict classicality for macrosystems} and {\em macrorealism} synonymously. (That said, I use the latter term only for historical reasons---it is misleading terminology, as I discuss in Section~\ref{realism}.)

However, this new definition has a number of advantages over earlier definitions. While Leggett and Garg's definition (Definition~\ref{LGdefn}) refers to an unspecified notion of state, the definition as strictly GPT classicality for macrosystems is formal and precise. Moreover, this definition makes explicit the fact that (as I argued in Section~\ref{mrps}) macrorealism as it is usually understood refers only to concepts that can be understood completely operationally---it is simply a hypothesis about what kind of operational theory describes one's degree of freedom. This is to be contrasted with other usages of the word `realism', which often invoke concepts (like hidden variables or causal structure) that are not defined in a direct operational manner. 
With respect to Timpson and Maroney's definition (Definition~\ref{opeigendefn}), my definition is the natural extension into the modern foundational landscape of operational theories. An advantage of this definition is that it does not specify in advance any macroscopic observable, whereas Timpson and Maroney's definition does refer to a given macroscopic observable (which in turn fixes the set of operational eigenstates and the dimensionality of the system). While one might have expected this difference to be inconsequential, we saw in Section~\ref{realizedNIM} (and we will see again in Section~\ref{error}) that it is not: making a priori assumptions about the scope of possible operational eigenstates can lead to loopholes in one's arguments. Additionally, my definition has the advantage that it specifies a class of full-bodied theories---e.g., including a specification of the measurements and transformations in the theory. One can then make use of all this structure (not merely facts about the states in the theory) in studying macrorealism. 

Moreover, the framework of generalized probabilistic theories is well developed, with a variety of tools and insights that we are now in a position to apply to the problem of testing macrorealism. For instance, in Section~\ref{strictertest}, this approach greatly clarifies the assumptions going into a new test of macrorealism, and in so doing helps us identify a serious flaw in the test.
Better still, in Section~\ref{tomog} we will see that an approach for characterizing GPTs known as theory-agnostic tomography provides a maximally informative and theory-independent test of whether or not a system is macrorealist. Finally, in Section~\ref{MRvsNC}, this new viewpoint will clarify the role that macrorealism plays as a notion of classicality---in particular, how it relates to the leading foundational notion of classicality, namely the existence of a generalized noncontextual ontological model. 

\section{Prior proposals for tests of macrorealism}

We now discuss three prior proposals for how macrorealism can be put to experimental test.

The most obvious way to test quantum behavior at a macroscopic level is to look for quantum interference with macrosystems. Leggett and Garg, however, were not satisfied by proposals of this sort. They noted that simply verifying some predictions of quantum theory (like interference phenomena) is not sufficient to rule {\em out} a competing class of theories (e.g., macrorealist ones), {\em unless} it has been shown that no theory in that competing class can reproduce the observed phenomena. Hence, they argued, what is needed is a theorem establishing that some set of experimental observations rules out macrorealism, analogous to how observed statistics which violate Bell inequalities can be used to rule out local causality. 

In the next three sections, I discuss three distinct proposals which aim to achieve this---to provide experimental means of falsifying macrorealism. My discussion of the first two of these is cursory, largely because they have already been criticized in prior works and in the discussion of realized-invasiveness.

\subsection{Leggett-Garg inequalities}

The first proposal for how one might test macrorealism was given by Leggett and Garg themselves. The relevant experiment, now known as the Leggett-Garg scenario, involves carrying out three different pairs of sequential measurements. Under the assumptions of macrorealism per se, noninvasive measurability, and realized noninvasiveness, Leggett and Garg proved that the correlations exhibited between the outcomes of these measurements are bounded by an inequality, known as the Leggett-Garg inequality. 
We refer the reader to Section 2-4 of Ref.~\cite{maroney2014quantum} for a clear and thorough discussion of the Leggett-Garg scenario and inequality.

Here, I simply wish to note that derivations of the Leggett-Garg inequality require the problematic assumption of realized noninvasiveness. It follows from the arguments in Section~\ref{realizedNIM} that the Leggett-Garg inequality is not capable of ruling out macrorealism without additional assumptions regarding the nature of the measuring device, and that such tests are consequently not as methodologically sound as, for example, tests of Bell inequalities or generalized noncontextuality inequalities.

\subsection{No-signaling in time}

The assumptions used by Leggett and Garg to derive their inequality are actually strong enough to generate contradictions with quantum theory in a dramatically broader class of scenarios than Leggett-Garg scenarios, and using arguments that are much more direct and elegant than the Leggett-Garg inequality.

In fact, the argument is quite obvious.
Provided that one grants Leggett and Garg's claim that a null-result measurement is nondisturbing when implemented on a macroscopic system in a macrorealist theory, it follows that one can falsify macrorealism simply by implementing a null-result measurement and then demonstrating that this measurement disturbs the statistics of any following measurement. For example, if the statistics of a measurement of the Pauli $Z$ operator are demonstrated to depend on whether or not a null-result measurement of the Pauli $X$ operator was done prior to it (as quantum theory predicts), then one has falsified Leggett and Garg's assumptions.

This argument was recognized in Refs.~\cite{Clifton1991,Grassi1994,GuidoLG,KoflerBrukner}. It seems to have been noted first by Clifton, who says
\begin{quote}
It is worth noting the simplicity of the above proof as against Leggett’s, which goes via a Bell-type inequality and uses a more complicated set of measurements. The motivation behind Leggett’s choice of proof is his desire to generate `the analog of what has been done in the EPR context: namely, to show that irrespective of the interpretational framework the experiments exclude a clearly formulated conjunction of ``common-sense'' hypotheses'. However the simple proof operates just like Leggett’s in not depending on QM’s formalism or its interpretation, and in only invoking its statistical predictions at the final state for comparison with macro-realist theories. So Leggett uses a sledgehammer to crack a nut.~\cite{Clifton1991}
\end{quote}
So even if one accepts Leggett and Garg's assumptions,  there is little reason to study Leggett-Garg inequalities in particular, since much simpler and more direct arguments suffice.

These ideas were expanded on and popularized by Kofler and Brukner~\cite{KoflerBrukner}, who proposed that one test Leggett and Garg's assumptions simply by checking whether or not a null-result measurement disturbs the statistics of any second measurement. They termed the macrorealistic prediction that there be no disturbance a {\em no-signaling in time} condition. 

Moreover, Kofler and Brukner recognized that standard demonstrations of interference are already enough to imply a violation of Leggett and Garg's assumptions. 
The argument is the same as before: provided that one grants Leggett and Garg's claim that a null-result measurement is nondisturbing when implemented on a macroscopic system in a macrorealist theory, it follows that one can use such a measurement to determine which way a particle went in an interferometer {\em without disturbing the interference pattern}. But this contradicts the predictions of quantum theory~\cite{Englert1996}.
So any experimental demonstration that a null-result measurement destroys interference fringes implies that one of Leggett and Garg's assumptions must be incorrect. Consequently, Kofler and Brukner's work shows that Leggett was too quick to believe~\cite{leggett2002testing} that observing quantum interference phenomena with macroscopic systems is insufficient to rule out macrorealism. 

Unfortunately, tests of no-signaling in time rely on realized noninvasiveness, and hence run into the same problems as tests of the Leggett-Garg inequality---they are just as theory- and model-dependent, and all the comments in Section~\ref{realizedNIM} apply.

\subsection{Nondisturbance conditions} \label{strictertest}

Recently, Leggett and coauthors introduced a new type of test~\cite{knee2016strict}   which does not rely on realized noninvasiveness. This test is essentially a synthesis of three ideas: Timpson and Maroney's more precise formulation of macroscopic realism as operational eigenstate mixture macrorealism~\cite{maroney2014quantum}, Kofler and Brukner's more general tests of macrorealism via no-signaling in time~\cite{KoflerBrukner}, and a paper by Wilde and Mizel which suggested that one could replace {\em assumptions} of noninvasiveness with {\em experimental quantification} of the amount of actual invasiveness~\cite{wilde2012addressing}. (Note that although the experiments and analyses in Ref.~\cite{knee2016strict} and Ref.~\cite{KoflerBrukner} are quite different, the nondisturbance conditions in Ref.~\cite{knee2016strict} are in spirit exactly the same kind of conditions as the no-signaling in time conditions in Ref.~\cite{KoflerBrukner}; I use different names for the two only as a manner of differentiating them, and in deference to the names chosen by their respective authors.) 

The logic of the argument is not particularly clear or explicit in Ref.~\cite{knee2016strict}. In making the argument more precise, I will show that it implicitly relies on an unphysical idealization that perfect pure states are prepared in one's experiment. When this idealization is dropped, one sees that the conditions derived in Ref.~\cite{knee2016strict} are no longer valid witnesses of the failure of macrorealism.

We wish to test the assumption of macrorealism, formalized as strict classicality, or equivalently (for the purposes of this argument), operational eigenstate mixture macrorealism.

Imagine that the macroscopic degree of freedom in question has $d$ operational eigenstates. As noted above, these correspond to vertices in a simplicial GPT, and I denote these by $\{{\bf s}_j\}_{j= 1,2,...,d}$. Now, imagine that one has found some transformation $\Phi$ and some measurement $M$ containing an effect ${\bf e}$ satisfying the constraint that for all $j$,
\begin{align} \label{controlexpnd}
{\bf e} \cdot \Phi {\bf s}_j = {\bf e} \cdot {\bf s}_j.
\end{align}
That is, the outcome statistics for the effect ${\bf e}$ on every given operational eigenstate are the same whether or not $\Phi$ is performed prior to the measurement.\footnote{Note that these conditions are similar to those for a transformation to be in the phase group of some GPT measurement~\cite{garner2013framework}.} As shorthand, we say that $\Phi$ does not disturb ${\bf e}$ (where the caveat that this need only be true for the operational eigenstates $\{ {\bf s}_j \}_j$ is left implicit). Then, it follows by linearity that $\Phi$ also does not disturb ${\bf e}$ for any GPT state in the convex hull of the operational eigenstates. That is, for ${\bf s} := \sum_j\alpha_j {\bf s}_j$ where $\alpha_j \in [0,1]$ and $\sum_j \alpha_j = 1$, one has
\begin{align} 
{\bf e} \cdot \Phi {\bf s} := {\bf e} \cdot \Phi \Big(\sum_j\alpha_j {\bf s}_j\Big)& = \sum_j\alpha_j {\bf e} \cdot \Phi {\bf s}_j \\ \nonumber
& = \sum_j\alpha_j {\bf e} \cdot {\bf s}_j = {\bf e} \cdot {\bf s}.
\end{align}
Thus, if one finds any state ${\bf v}$ for which $\Phi$ {\em does} disturb the outcome statistics of ${\bf e}$, so that 
\begin{equation} \label{dist}
{\bf e} \cdot \Phi {\bf v} \neq {\bf e} \cdot {\bf v},
\end{equation}
then it follows that ${\bf v}$ is not a convex combination of the operational eigenstates.

The intuition behind the proposal of Ref.~\cite{knee2016strict} is now straightforward.
One begins by implementing a control experiment, wherein one verifies that Eq.~\eqref{controlexpnd} holds for the operational eigenstates (for some measurement M and some transformation $\Phi$). Then, one demonstrates the existence of any other state ${\bf v}$ for which Eq.~\eqref{dist} holds.
Hence, ${\bf v}$ is not in the convex hull of the operational eigenstates. Assuming that the true set of operational eigenstates for the system in question is indeed given by $\{{\bf s}_j\}_j$, one can conclude that there exist states outside the simplex defined by their convex hull. In other words, the GPT's state space is strictly larger than the simplicial GPT containing them. So, if such a state ${\bf v}$ can be found, macrorealism is falsified.

\begin{figure}[htbp] 
   \centering
   \includegraphics[width=1\columnwidth]{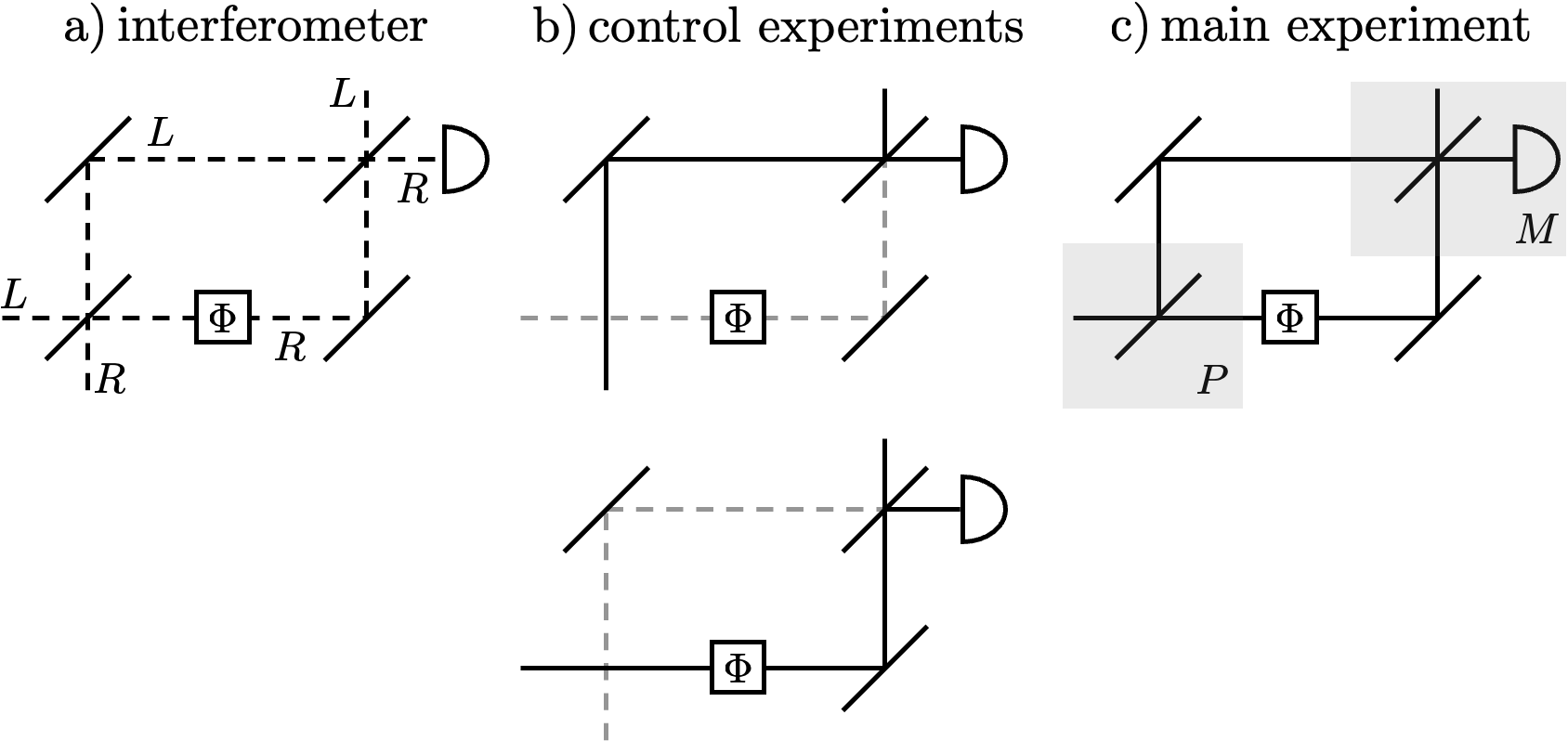} 
   \caption{Recasting an interferometric experiment as a test of the nondisturbance condition in Ref.~\cite{knee2016strict}.}
   \label{interf}
\end{figure}

A standard interferometer can be recast into the mold of this experimental proposal. (This is not necessary for the argument, but it is insightful.) Consider an interferometer with two paths, $L$ and $R$, as depicted in Fig.~\ref{interf}(a). The measurement $M$ is a measurement of which output port the particle takes after being passed through the second beamsplitter, as depicted by the shaded box labeled $M$ in Fig.~\ref{interf}(c), and the effect ${\bf e}$ can be taken to be the case where the particle takes output $R$. (This measurement is sometimes termed a which-phase measurement, since its two eigenstates correspond to a relative phase of $0$ and of $\pi$ between the two paths.) One assumes that there are two operational eigenstates in the experiment---the which-path eigenstates---and the operation $\Phi$ is taken to be a phase shift of $\pi$ in path $R$. When the operational eigenstate corresponding to the particle traversing path $R$ is prepared, as shown in the top of Fig.~\ref{interf}(b), the phase shifter $\Phi$ does not disturb the statistics of the measurement (whose outcome is uniformly random, since there is no interference, and the particle is equally likely to take either output port). The same is true when the operational eigenstate corresponding to the particle traversing path $L$ is prepared, as shown in the bottom of Fig.~\ref{interf}(b). These two cases constitute the control experiment. In the final stage of the experiment, one demonstrates the existence of a state ${\bf v}$ for which the phase shifter {\em does} affect the statistics of the measurement. Assuming the validity of quantum theory, one such state is given by a coherent superposition of the two paths, e.g., the state prepared by the shaded box labeled $P$ in Fig.~\ref{interf}(c).

For simplicity, I have given the argument so far under an idealization, which we must now relax. In a real experiment, there will be nonzero disturbance even in the control experiment, so that Eq.~\eqref{controlexpnd} is not perfectly satisfied. To deal with this, one simply quantifies the disturbance for each ${\bf s_j}$ via\footnote{Note that Ref.~\cite{knee2016strict} actually works with expectation values rather than probabilities for individual effects, and hence quantifies disturbance in a slightly different manner. The two approaches are equivalent up to algebra.}
\begin{equation} 
d_j:=|{\bf e} \cdot \Phi {\bf s}_j -{\bf e} \cdot {\bf s}_j|.
\end{equation}
It is not hard to see that the disturbance to any state in the convex hull of the ${\bf s_j}$ must then be less than the maximum disturbance $\max\limits_j[d_j]$. 
Explicitly, consider again the convex mixture ${\bf s}:=\sum_j \alpha_j {\bf s}_j$ defined above, and note that the disturbance $d_s$ to this state is given by 
\begin{align}
d_s &= |{\bf e} \cdot \Phi {\bf s} -{\bf e} \cdot {\bf s}|\\
&=|{\bf e} \cdot \Phi \Big(\sum_j \alpha_j {\bf s}_j\Big)-{\bf e} \cdot \Big(\sum_j \alpha_j {\bf s}_j\Big)| \\
&=|\sum_j \alpha_j \Big({\bf e} \cdot \Phi {\bf s}_j-{\bf e} \cdot {\bf s}_j\Big)| \\
&\leq\sum_j |\alpha_j \Big({\bf e} \cdot \Phi {\bf s}_j-{\bf e} \cdot {\bf s}_j\Big)| \label{pos1} \\
&=\sum_j \alpha_j |\Big({\bf e} \cdot \Phi {\bf s}_j-{\bf e} \cdot {\bf s}_j\Big)| \\
&=\sum_j \alpha_j d_j \\
&\leq \sum_j \alpha_j \max\limits_{j'}[d_{j'}] \label{pos2}\\
&= \max\limits_{j'}[d_{j'}] \sum_j \alpha_j \\
&=\max\limits_{j'}[d_{j'}]. \label{pos3}
\end{align}
Note that Eq.~\eqref{pos1} and Eq.~\eqref{pos2} only hold because $\alpha_j\ge0$ for all $j$, and Eq.~\eqref{pos3} holds because $\sum_j \alpha_j = 1$. (Hence, this logic does not go through for states in the linear span of the operational eigenstates but which are not in the convex hull.)

Consequently, Ref.~\cite{knee2016strict} proposes that one can falsify macrorealism  by demonstrating the existence of any state ${\bf v}$ for which the disturbance is greater than any of the disturbances in the control experiment---that is, greater than $\max\limits_j[d_j]$. (And as we saw in the interferometer example above, quantum theory predicts that such states can easily be found.) In such a case, we can conclude that ${\bf v}$ is outside the convex hull of the states $\{ {\bf s}_j \}_j$ implemented in one's control experiment. Assuming that the $\{ {\bf s}_j \}_j$ are in fact the true operational eigenstates, it follows that ${\bf v}$ is outside the convex hull of the operational eigenstates, and consequently the system cannot be macrorealist.

Unfortunately, in any real control experiment, it will never be the case that the states $\{ {\bf s}_j \}_j$ are exactly the operational eigenstates, and so this last step in the argument fails.

\subsubsection{An error in the argument} \label{error}

What Ref.~\cite{knee2016strict} genuinely does establish is that there exist states outside the convex hull of the states $\{ {\bf s}_j \}_j$ actually realized in one's control experiment. But Ref.~\cite{knee2016strict} implicitly assumes that the $\{ {\bf s}_j \}_j$ are exactly the operational eigenstates (and moreover that these states include {\em all possible} operational eigenstates), and this is never the case. 
That is, the actual preparations realized in one's control experiment are never perfect, but rather are always noisy. In a macrorealist theory in particular, they necessarily correspond to {\em nontrivial mixtures} of operational eigenstates. Consequently, the states $\{ {\bf s}_j \}_j$ and their convex hull will be {\em strictly} contained in the full simplex of states in a macrorealist theory. 
So, one cannot rule out macrorealism by establishing that there exist states outside the convex hull of the $\{ {\bf s}_j \}_j$, since states lying outside this convex hull but inside the full simplex are still consistent with macrorealism.

\begin{figure}[htbp] 
   \centering
   \includegraphics[width=.99\columnwidth]{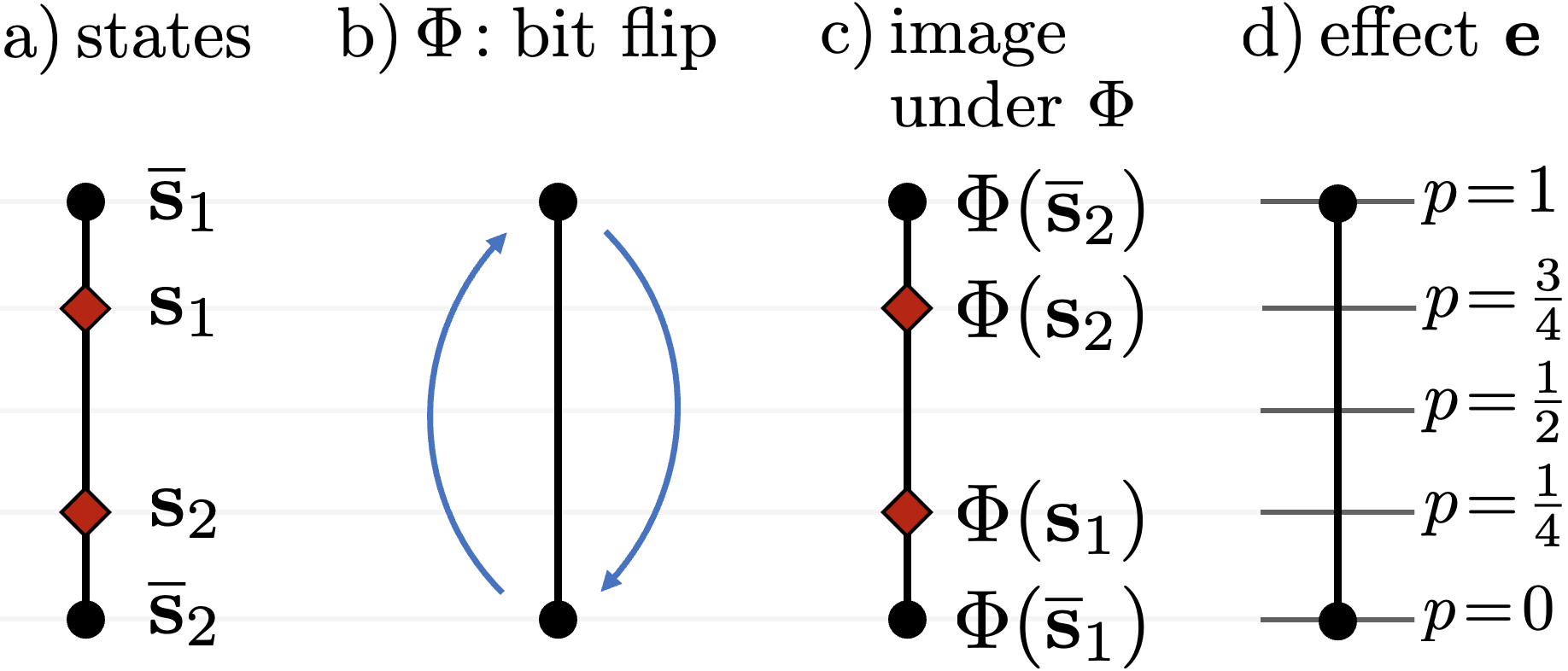} 
   \caption{ Consider a macrorealist system with two operational eigenstates $\overline{{\bf s}}_1$ and $\overline{{\bf s}}_2$, and imagine that when one tries to prepare these in one's control experiment, one actually prepares the mixed states ${\bf s}_1$ and ${\bf s}_2$. If one takes $\Phi$ to be a bit flip and takes ${\bf e}$ to be the effect associated to the $\overline{{\bf s}}_1$ outcome of the measurement that perfectly discriminates $\overline{{\bf s}}_1$ and $\overline{{\bf s}}_2$, then the analysis of Ref.~\cite{knee2016strict} leads to the incorrect conclusion that one has falsified macrorealism.}
   \label{example1}
\end{figure}

A concrete example, depicted in Figure~\ref{example1}, illustrates this. Take the macrosystem to possess two operational eigenstates, which I denote by $\overline{{\bf s}}_1$ and $\overline{{\bf s}}_2$. Such a system is described by a classical bit. The corresponding simplex of mixed GPT states, depicted in Figure~\ref{example1}(a), is simply the line parametrized by the mixed states $\alpha \overline{{\bf s}}_1 + (1-\alpha)\overline{{\bf s}}_2$ as $\alpha$ runs from $0$ to $1$. Moreover imagine that when one attempts to prepare the extremal states $\overline{{\bf s}}_1$ and $\overline{{\bf s}}_2$, one's apparatus actually prepares noisy versions of these described by the mixed states ${\bf s}_1=\frac34 \overline{{\bf s}}_1 + \frac14 \overline{{\bf s}}_2$ and ${\bf s}_2=\frac14 \overline{{\bf s}}_1 + \frac34 \overline{{\bf s}}_2$.
We take the operation $\Phi$ to be the linear map on the state space induced by swapping $\overline{{\bf s}}_1$ and $\overline{{\bf s}}_2$, as depicted in Figure~\ref{example1}(b). The image of the relevant states under this transformation is depicted in Figure~\ref{example1}(c). Finally, imagine that the final measurement $M$ is simply a measurement of which operational eigenstate the system is in, and focus on the outcome corresponding to $\overline{{\bf s}}_1$. In Figure~\ref{example1}(d), I plot (as level curves) the probabilities induced on the state space by this GPT effect.
(This example is somewhat exaggerated, in the sense that the noisy versions of the operational eigenstates are far from the operational eigenstates themselves, and the process $\Phi$ is far from being nondisturbing. But it makes the point cleanly.)

One can read off the disturbance induced by $\Phi$ to the effect in question (for each given state) from Figure~\ref{example1}:
\begin{align}
d_{{\bf \overline{s}}_1}&= |1-0|=1\\
d_{{\bf \overline{s}}_2}&= |0-1|=1\\
d_{{\bf s}_1}&= |\frac34 - \frac14| = \frac12\\
d_{{\bf s}_2}&= |\frac14 - \frac34| = \frac12.
\end{align}
In the analysis of Ref.~\cite{knee2016strict}, one would conclude from the observed control disturbances $d_{{\bf s}_1}=d_{{\bf s}_2}=\frac12$ that the disturbance for all possible states in a macrorealist theory must be less than $\frac12$. But we see explicitly in this example that there exist states (e.g., the true operational eigenstates) of the system in question for which the disturbance can be greater than $\frac12$, despite the fact that the system is explicitly macrorealist.

So one cannot falsify macrorealism simply by finding a state for which the disturbance is greater than the measured value $\max\limits_j[d_j]$.

There is a second problem with the argument as well. As mentioned earlier, a second implicit assumption required by the proposal in Ref.~\cite{knee2016strict} is that the set of states in one's control experiment includes {\em all possible} operational eigenstates. Imagine that the true degree of freedom being tested has three possible operational eigenstates, but that one is not aware of this fact, but rather only recognizes the existence of two of these. Consequently, one would only have quantified the disturbance on the two known operational eigenstates. It is entirely possible in such a case that the disturbance on the third operational eigenstate is higher than the disturbance on the two for which the disturbance was characterized, in which case one's experimental estimation of $\max\limits_j[d_j]$ would be mistaken, and consequently so too could one's assessment of macrorealism. 

\begin{figure}[htbp] 
   \centering
   \includegraphics[width=.7\columnwidth]{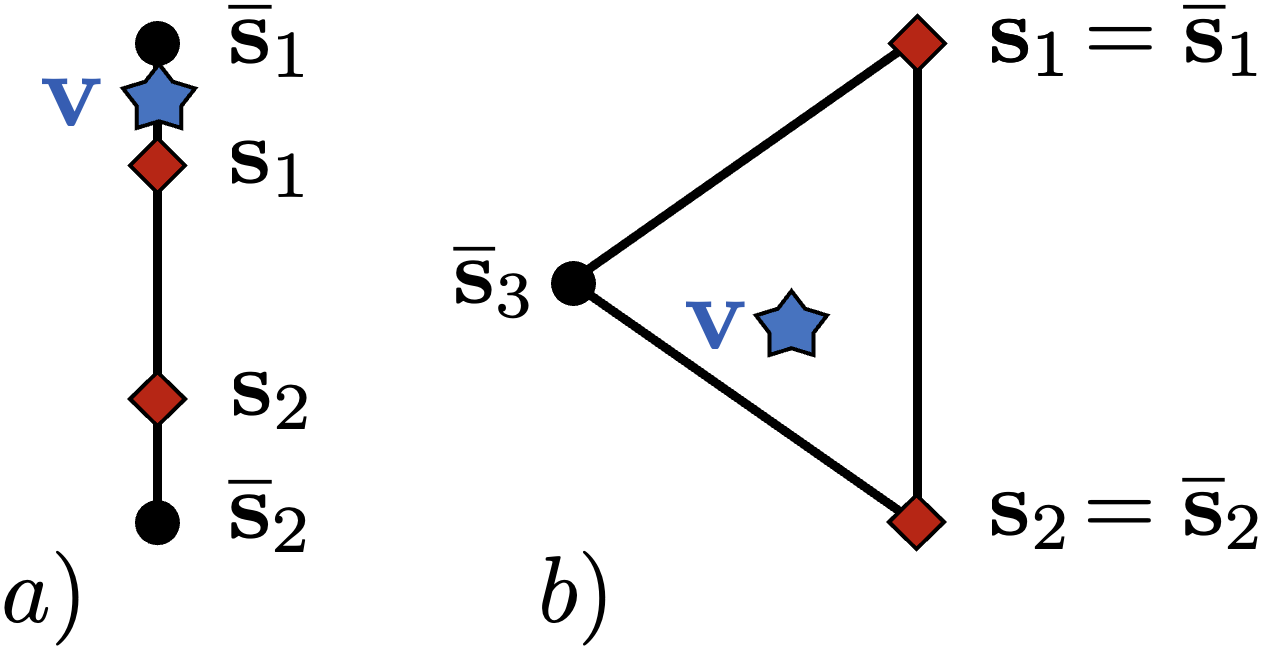} 
   \caption{
Two distinct ways in which one can find a state ${\bf v}$ which is {\em outside} the convex hull of the states ${\bf s}_1$ and ${\bf s}_2$ realized in one's control experiment, while still being {\em inside} the convex hull of the true operational eigenstates (the $\overline{{\bf s}}_j$). In a), this possibility is a consequence of noise; in b), it is a consequence of there being a third macrostate $\overline{{\bf s}}_3$ which was not accounted for in the control experiment.
}
   \label{twowaysout}
\end{figure}

In summary, what the test proposed in Ref.~\cite{knee2016strict} claims to demonstrate is that there exists a state ${\bf v}$ outside the convex hull of the true operational eigenstates. But what it really demonstratesis that there exists a state ${\bf v}$ outside the convex hull {\em of the states realized in one's control experiments}, and this is possible even in a macrorealist theory.
In fact, we have seen two distinct ways that this may happen (depending on whether or not ${\bf v}$ lies in the span of the actually realized ${\bf s}_1$ and ${\bf s}_2$) ; I depict these graphically in Fig.~\ref{twowaysout}.

\subsubsection{Patching up the error?} \label{patch}

We have just shown how tests of the nondisturbance condition rely on idealizations regarding the states realized in one's control experiment: that one is aware of all possible operational eigenstates, and moreover that one has perfectly prepared each of them in the control experiment. Even if one grants the first of these, the latter is {\em never} satisfied in a real experiment, and consequently the witness derived in Ref.~\cite{knee2016strict} is not robust to noise.
However, the idealized argument is sound, and so one might naturally wonder if this problem cannot be patched up simply by carrying out a more careful analysis and deriving a noise-robust witness. 

To begin with, one could assume that the true set of operational eigenstates is known for the macroscopic observable of interest. Recall that this is exactly the same theory-dependent assumption that was implicitly needed to justify the assumption of realized noninvasiveness used in prior proposals for falsifying macrorealism. As noted in Section~\ref{realizedNIM}, one might presuppose that there are exactly two macrostates for the system of interest: `located at the right slit and `located at the left slit'. But as also discussed previously, additional assumptions like this diminish the scope of theories that can be falsified by such tests, and deviate from the theory-independent ideals that one should aim to meet (and that Leggett and Garg themselves aimed to meet).

One could then introduce additional measurements to the scenario in an attempt to characterize how close one's control preparations are to being perfect operational eigenstates. However, to have any confidence in this characterization, one would require access to a well-characterized set of measurements. But given that the aim is to make no a priori assumptions about the nature of one's preparations and measurements, this is not a valid option.
So, it seems we are bound by a catch-22: how could one characterize a set of states without prior knowledge of one's measurements, if this prior knowledge can only be obtained by implementing the measurements on a well-characterized set of states?

As it turns out, prior work in the framework of GPTs provides us exactly the tool we need to avoid this circularity.

\section{A maximally informative and theory-independent test of macrorealism} \label{tomog}

Let us take stock of the proposed tests we have discussed so far. Tests based on the Leggett-Garg inequality and on no-signaling in time rely on an idealization regarding the {\em measurements} realized in one's experiment: that these are perfectly nondisturbing. Arguments which attempt to justify perfect nondisturbance rely on theory-dependent assumptions regarding the specific physics of the interaction realizing one's measurement and regarding the specific sorts of properties that are relevant for explaining one's observations.
Meanwhile, tests of the nondisturbance condition rely on idealizations regarding the {\em states} realized in one's experiment: that one is aware of all possible operational eigenstates, and moreover can perfectly prepare each of them. In order to make the argument robust to these idealizations, one {\em again} requires theory-dependent assumptions, in particular, that one has well-characterized measurements with which to obtain a characterization of one's control preparations.

Thus, while all of these tests may teach us something, they are not methodologically on par with state-of-the-art foundational experiments such as Bell tests or tests of generalized noncontextuality. 

However, once one recognizes that macrorealism corresponds to an entirely operational hypothesis about the nature of the GPT governing one's macroscopic degree of freedom, one can see immediately that an existing method for experimentally determining the GPT governing a given system provides a direct means of testing macrorealism (when it is applied to a macrosystem). This method is known as bootstrap tomography or {\em theory-agnostic tomography}~\cite{bootstrap1,bootstrap2}. 

In theory-agnostic tomography, one performs a large number of preparations and measurements on the system in question and records the statistics that arise. These procedures may be chosen to target specific states and effects, but they may just as well be chosen randomly. In fact, one does not assume anything a priori about the nature or identity of each individual preparation or measurement, nor about the true GPT governing the experiment. From the data one collects, one then carries out an analysis which extracts this information {\em as output}---namely, one finds the GPT descriptions of every state and effect realized in one's experiment. It follows that the convex hull of these realized state vectors provides an (inner) approximation of the true GPT state space, and the convex hull of the realized effect vectors provides an (inner) approximation of the true GPT effect space.\footnote{Note that one can moreover compute {\em outer} approximations of the true GPT from these.} I will refer to these inner approximations as the state and effect spaces of the {\em realized GPT}. (The accuracy of these approximations, of course, depends on the choice and number of preparations and measurements, as well as how noisy these are. But they are necessarily inner bounds, and this is all that matters for our purposes.)

Once one has obtained the realized GPT describing some macrosystem, then, it is simple to determine if the experiment falsifies macrorealism: one must simply check whether the realized GPT can be embedded into the unique strictly classical GPT of the same dimension. More precisely, the experiment is consistent with macrorealism if and only if there is a linear embedding of the realized GPT state space into a simplex of the same dimension and a linear embedding of the realized GPT effect space into the logical dual to this simplex, such that the embeddings generate the same probabilities as the realized GPT itself. If no such embedding is possible, then one has falsified macrorealism. 

Note that one can never {\em verify} macrorealism, since any real experiment will not fully explore the true GPT state space and effect space. So if a simplicial embedding does exist, the most one can say is that the experiment is {\em consistent with} macrorealism. However, one can gather {\em evidence} that the system is macrorealist by showing that the realized state and effect space {\em approach} a simplex and its dual as more (and less noisy) preparations and measurements are performed on the system.

There are several key advantages of this approach to testing macrorealism.

Theory-agnostic tomography provides more information, and more direct information, than prior tests of macrorealism, which tested specific witnesses of the failure of macrorealism. For example, the approach of Ref.~\cite{knee2016strict} aims to establish that there is a single state living outside the true simplex of states. But in the approach of theory-agnostic tomography, one can probe the full convex geometry of the state and effect spaces, including {\em all possible states} lying outside the simplex (not to mention a characterization of the effect space as well). 
One could even extend this approach to characterize the transformative and compositional aspects of the system of interest, and use all this additional information to determine consistency with macrorealism. (However, this extension remains to be worked out explicitly, since prior proposals for, and demonstrations of, theory-agnostic tomography have only been given for prepare-measure experiments.)

Additionally, this technique is theory-independent, in that it does not assume that one has well-characterized preparations or measurements at the start of the experiment, nor must one assume anything about the theory governing the experiment. 
The primary assumption going into this procedure is that the sets of realized preparations and measurements contain tomographically complete states and effects. In other words, theory-agnostic tomography relies on the assumption that one's laboratory operations implement a set of GPT states and GPT effects that span the true GPT state and effect spaces\footnote{In fact, slightly weaker assumptions sometimes suffice~\cite{labnotebook,PuseydelRio}.}. Note that one need not know or assume anything a priori about which particular preparations and measurements form the tomographically complete sets, nor must the preparations and measurements satisfy any other idealizations (like extremality or nondisturbance). As such, this assumption is weaker than those used in Ref.~\cite{knee2016strict}, which required not only that one's realized control states spanned the true simplex of states, but {\em additionally} that these realized control states be perfectly extremal (pure).
Moreover, one can gather {\em experimental evidence} for tomographic completeness, as was discussed in detail in Ref.~\cite{bootstrap1}.  

A final benefit of the fact that this procedure gives a complete GPT characterization of one's system is that one can use this same data to falsify not only macrorealism, but also the much broader class of theories defined by a more permissive notion of classicality (namely, generalized noncontextuality). I return to this point in Section~\ref{MRvsNC}.

In summary, carrying out theory-agnostic tomography on a macroscopic system provides a maximally informative method of falsifying macrorealism, and one that moreover needs none of the theory-dependent assumptions required in prior tests of macrorealism.

\section{Discussion}

\subsection{Is macrorealism a notion of realism?} \label{realism}

I now argue that macrorealism is {\em not} a principle characterizing scientific realism or any class of `realist theories'. 

This view is shared by many earlier authors.
Benatti, Ghirardi, and Grassi~\cite{Grassi1994} argue that noninvasive measurability {\em ``is by no means necessary for realism but it embodies classical demands''} and that violation of the Leggett-Garg inequality {\em``would not imply the impossibility of a realistic position about the property that the current is going left or right, but the impossibility of having a theory exhibiting, at the individual level, classical features''}. Timpson and Maroney~\cite{maroney2014quantum} provide detailed arguments for why ``{\em macroscopic realism is not equivalent to realism about the macroscopic}'', and that, Leggett and Garg's arguments notwithstanding,``{\em there is nothing realist about denying the existence of superpositions, macroscopic or otherwise}''. Bacciagaluppi~\cite{GuidoLG} also grants that violation of the Leggett-Garg inequality may be a signature of nonclassicality, but argues that it does not rule out the possibility of a realist understanding of macroscopic systems, with Bohmian mechanics as his counterexample.

Strictly speaking, realism is not a property of a physical theory, but rather is a kind of attitude towards a physical theory (or more broadly, towards the practice of science itself). In particular, it is the view that our best physical theories give or aim to give a literally true description of the world. To take a realist attitude towards a particular theory is to believe the statements it makes (or at least some substantial subset thereof) to be true, and to believe that the concepts within the theory (or at least some substantial subset thereof) correspond directly to entities in the world. 

So it is an abuse of terminology to state that a class of theories `is realist'. Still, there is by now a fairly standard understanding of what is meant by such a statement: that the class of theories in question can be easily interpreted in a realist manner---i.e., without requiring a radical change to our traditional, classical notions of kinematical state spaces and dynamics. For instance, there is little obstacle to taking Newtonian concepts such as forces and masses to be literal descriptions of entities in the world, but the problem of associating quantum concepts such as wavefunction amplitudes or unitaries to entities in the world is much more contentious.

With this understanding of the terminology, one can view macrorealism as a {\em sufficient} condition for the operational description of a macroscopic degree of freedom to `be realist'---that is, to admit of a natural realist interpretation. As I discussed just after defining them, any strictly simplicial GPT system is in one-to-one correspondence with a classical random variable whose values can be taken as the possible macroscopic properties of the system in question. 

However, strict GPT classicality is not a {\em necessary} condition for a theory to admit of a realist interpretation---indeed, macrorealist theories are a set of measure zero in the space of theories that can be viewed as realist. Even if one focuses on operational theories (i.e., GPTs), the much broader class of simplex-embeddable theories admit of a realist interpretation in exactly the same way that strictly classical theories do~\cite{SchmidGPT} (but where one additionally imposes a limit on what can be known about the true state of the system~\cite{spekkens2007evidence,epistricted,bartlett2012reconstruction,catani2021interference}). Moreover, one can be realist about theories that are not  operational, such as Bohmian mechanics. And because Bohmian mechanics reproduces quantum theory, it follows that one can maintain realism (about both the macroscopic {\em and} the microscopic) for all of quantum theory. Indeed, it is hard to say what it would even mean for a theory to `{\em not} admit of any realist interpretation'---it is just that the precise manner in which one upholds one's realist ideals tends to become more nuanced as one considers more exotic theories.

In summary, the term macrorealism is poor terminology. The notion in question has little to do with `a class of theories that can be viewed as providing a realist description of macroscopic objects', as its name would suggest. Rather, it singles out a set of theories that are classical (for macroscopic properties) in the strongest sense of the word. As such, I advocate that the term macrorealism be replaced by {\em strict classicality for macrosystems} (or strict simpliciality for macrosystems).

\subsection{Macrorealism versus generalized noncontextuality as notions of classicality} \label{MRvsNC}

The question of which theories admit of a classical explanation has been the subject of a great deal of research. 
In my view, the most interesting notion of classical-explainability is that of generalized noncontextuality. In this section, I briefly motivate this claim, and then I discuss how strictly classical theories are a special case of generalized noncontextual theories. I will focus on the prepare-and-measure case, although most of what I say can be extended naturally to arbitrary circuits~\cite{schmid2020structure}. 

The considerations that follow are true for microsystems as well as macrosystems, so I will focus throughout on strict classicality rather than macrorealism. The connection to macrorealism is (as always) made simply by considering the special case where these considerations are applied to macroscopic systems in particular.

The list of arguments in favor of taking generalized noncontextuality as one's notion of classicality is long and growing longer. It can be motivated by a version of Leibniz's principle of the identity of indiscernibles, a principle for theory construction that has been used to great success in physics in the past~\cite{Leibniz}. It is equivalent to a notion of classicality used within quantum optics (namely, the existence of a positive quasiprobability representation~\cite{negativity,SchmidGPT,schmid2020structure}) and to a notion of classicality within the framework of generalized probabilistic theories (namely, being a subtheory of a simplicial GPT~\cite{SchmidGPT}). Additionally, noncontextuality emerges in the classical limit considered within the quantum Darwinist research program~\cite{baldijao2021noncontextuality} and under sufficient noise or coarse-graining~\cite{marvian2020inaccessible,selby2022open}. Furthermore, other key indicators of nonclassicality, such as violations of local causality~\cite{Bell} or observations of anomalous weak values~\cite{AWV} imply the failure of generalized noncontextuality. Finally, contextuality is a resource for information-processing~\cite{POM,RAC,RAC2,Saha_2019}, computation~\cite{comp1,comp2,schmid2021only}, state discrimination~\cite{schmid2018contextual}, cloning~\cite{cloningcontext}, and metrology~\cite{PhysRevLett.125.230603}. 

Although generalized noncontextuality was originally defined in the context of unquotiented operational theories and ontological models~\cite{gencontext}, we will not need this definition here. Operational theories that admit of a generalized noncontextual model correspond to GPTs which satisfy a property known as simplex-embeddability~\cite{SchmidGPT,selby2021accessible}, and we can take this as our characterization of noncontextuality in this work.
A prepare-measure GPT is simplex-embeddable if and only if it can be linearly mapped into a strictly classical GPT (possibly of higher dimension than the given GPT's dimension) in such a way that the probabilities predicted by the theory are preserved. For the reasons listed in the previous paragraph, such GPTs are said to be {\em classically-explainable}. 

The relationship between strict classicality and generalized noncontextuality is now seen to be quite simple. Strictly classical GPTs are a special case of noncontextual theories, since a simplicial theory is trivially embeddable in a simplicial theory, namely itself. Indeed, simplicial theories are all and only~\cite{SchmidGPT,shahandeh2019contextuality} those simplex-embeddable theories with no epistemic restriction~\cite{spekkens2007evidence}. Geometrically, the simplicial theories form a set of measure zero in the set of all simplex-embeddable theories, so strict classicality (and consequently macrorealism as well) is a dramatically more restrictive notion of classicality than generalized noncontextuality.

Generalized noncontextuality provides a nuanced dividing line between those phenomena and theories that are classical and those that are nonclassical. Many of the qualitative quantum phenomena that have been thought genuinely nonclassical can in fact be recovered in noncontextual theories. This is best seen by looking at those phenomena reproduced in simplex-embeddable theories that are not strictly classical (often known as epistemically restricted classical theories~\cite{spekkens2007evidence,epistricted,bartlett2012reconstruction,catani2021interference}). For instance, in such theories, teleportation is possible, an unknown state cannot be cloned, interference patterns can be observed, there exist entangled states, and so on. I refer the interested reader to Ref.~\cite{epistricted} (and in particular, Table~II therein) for more details. Most interesting is the set of phenomena which {\em cannot} be reproduced in any noncontextual theory, such as violations of Bell inequalities and computational speedups. Taking noncontextuality as our notion of classicality, then, implies that it is these latter sorts of phenomena that one should focus on when trying to understand the essence of quantum theory (rather than phenomena like interference, entanglement, and so on), and that one should expect to be useful as quantum resources~\cite{Barrett2005PRresource,Asymy1,veitch2014resource,coecke2016mathematical,Rosset-Buscemi-Liang-PRX,Asymy2,LOSRvsLOCCentang,Schmid2020typeindependent}. 

Strict classicality, and consequently macrorealism, defines a different dividing line between classical and nonclassical---one for which a much wider class of phenomena are deemed nonclassical (including interference, entanglement, and so on). in Ref.~\cite{epistricted}, such phenomena were termed `weakly nonclassical'.

Which notion of classicality is more interesting in the context of studying the behavior of macroscopic systems may depend on one's prior expectations for such systems.
If one's goal is to demonstrate {\em nonclassicality} (as is typically the case, e.g., in experiments aiming to witness interference with macrosystems), then it is more interesting to demonstrate the failure of simplex-embeddability than the failure of simpliciality, since this is a stronger witness of nonclassicality. However, if one's goal is to demonstrate {\em classicality} (which might be a natural goal for someone---perhaps a collapse theorist---with the a priori expectation that macroscopic systems are fundamentally classical rather than quantum), then it might be more interesting to show that a given system is simplicial in addition to being simplex-embeddable. For if one can demonstrate that a degree of freedom is not only noncontextual but moreover strictly classical, then one can be assured that the true state of this system can be perfectly known, cloned, measured without disturbance, and so on. 

Finally, I expand on how performing theory-agnostic tomography on a macroscopic system allows one to simultaneously test for macrorealism and for generalized noncontextuality. Recall from Section~\ref{tomog} that theory-agnostic tomography allows one to compute an inner approximation (termed the realized GPT) of the true GPT describing one's experiment. Recall also that a macrosystem is consistent with macrorealism if and only if the realized state and effect spaces embed into a simplex of the same dimension and its dual, respectively.
In contrast, to decide if a system is classically-explainable, one must determine if its state and effect space embed in a simplex {\em of arbitrary dimension} and its dual. While this sounds more difficult to compute, it was shown in Ref.~\cite{selby2022open}, that such an embedding exists if and only if an embedding exists in the square of the given GPT's dimension, and moreover that one can settle this question with a linear program. If one finds that the realized GPT states and effects do not admit of a simplex-embedding of this relaxed sort, then one has witnessed the nonclassicality of the system in question.

Hence, for any given macrosystem, there are three possibilities. First, if the realized GPT embeds in a simplicial GPT of dimension $d$, then the experiment is consistent with both noncontextuality and macrorealism. (As before, one can never prove that the system being probed {\em is} macrorealist or noncontextual, since any real experiment will not fully explore the true GPT state space and effect space. As with any scientific hypothesis, one can only gather evidence to support these hypotheses, or falsify them.)
Second, if the realized GPT embeds in a simplicial GPT of dimension $d^2$ but {\em not} in a simplicial GPT of dimension $d$, then the experiment admits of a noncontextual explanation but not a macrorealist one. 
Finally, if the realized GPT is not embeddable in a simplicial GPT of dimension $d^2$, then the experiment has witnessed the failure of both macrorealism and noncontextuality.

\section*{Acknowledgments}

I thank Haoxing Du, John Selby, and Rob Spekkens for useful discussions and comments on the draft. I acknowledge support by the Foundation for Polish Science (IRAP project, ICTQT, contract no.2018/MAB/5, co-financed by EU within Smart Growth Operational Programme). 

\bibliographystyle{quantum}
\bibliography{bib.bib}

\end{document}